\documentclass[11pt]{amsart}
\usepackage[a4paper,margin=2.8cm]{geometry}
\usepackage{amsmath,amssymb,amsthm}
\usepackage{url}
\usepackage[hidelinks]{hyperref}
\hypersetup{pdftitle={The Bapat--Sunder eigenvalue conjecture fails over the reals},pdfauthor={Logan R. Chalmers}}
\newtheorem{theorem}{Theorem}
\newtheorem{lemma}{Lemma}
\newtheorem{corollary}{Corollary}
\newtheorem{conjecture}{Conjecture}
\newcommand{\per}{\operatorname{per}}
\newcommand{\R}{\mathbb{R}}
\newcommand{\Z}{\mathbb{Z}}

\title{The Bapat--Sunder eigenvalue conjecture fails over the reals}
\author{Logan R. Chalmers}
\address{University of Otago, Dunedin 9016, New Zealand}
\email{logan.chalmers@postgrad.otago.ac.nz}
\thanks{ORCID: 0009-0001-6468-3889.}
\date{}

\begin{document}
\begin{abstract}
In 1986 Bapat and Sunder asked whether $\per(A)$ is the largest
eigenvalue of $F_A=[a_{i,j}\per(A(i,j))]$ for every positive
semidefinite Hermitian $A$, where $A(i,j)$ deletes row $i$ and column
$j$. Drury disproved this over the complex field in 2018 and
conjectured that the property holds for real positive semidefinite
matrices. We disprove Drury's conjecture with a rank-$4$, $32\times32$
integer Gram matrix and an integer vector whose Rayleigh quotient
exceeds $\per(A)$; verification uses integer arithmetic.
Together, the two counterexamples resolve the Bapat--Sunder eigenvalue
conjecture negatively over both fields. Counterexamples exist of every
order $n\ge32$. As an application to quantum optics, the
violation yields a $36$-mode orthogonal interferometer and a direction
of relative arrival-time delays among $32$ photons with
identical-variance Gaussian wave packets along which every sufficiently
small nonzero delay increases the probability that all photons are
detected within four designated output modes.
\end{abstract}
\maketitle

\section{Introduction}

The permanent of an $n\times n$ matrix $A=[a_{i,j}]$ is
\begin{equation}
\per(A)=\sum_{\sigma\in S_n}\prod_{i=1}^{n}a_{i,\sigma(i)},
\end{equation}
where $S_n$ denotes the symmetric group. For positive semidefinite $A$
one has $\per(A)\ge 0$ \cite{Minc}. Given a positive semidefinite
Hermitian $A$, define the $n\times n$ matrix $F_A=[f_{i,j}]$ by
\begin{equation}\label{eq:FA}
f_{i,j}=a_{i,j}\,\per\bigl(A(i,j)\bigr),
\end{equation}
where $A(i,j)$ is the $(n-1)\times(n-1)$ submatrix of $A$ obtained by
deleting the $i$-th row and the $j$-th column; we call $F_A$ the
\emph{Bapat--Sunder matrix} of $A$. The Laplace expansion of
the permanent gives $\sum_j f_{i,j}=\per(A)$ for every $i$, so
$\per(A)$ is an eigenvalue of $F_A$ with the all-ones eigenvector;
moreover $F_A$ is again positive semidefinite \cite{BapatSunder1986,
Drury2018}. When $A$ is real, $F_A$ is moreover real symmetric: since
$A(i,j)^{\mathsf T}=A(j,i)$ and $\per(M^{\mathsf T})=\per(M)$, we have
$f_{j,i}=a_{j,i}\per(A(j,i))=a_{i,j}\per(A(i,j))=f_{i,j}$.

In 1986 Bapat and Sunder \cite{BapatSunder1986} asked whether $\per(A)$
is necessarily the \emph{largest} eigenvalue of $F_A$; since $\per(A)$
is always an eigenvalue, the question is equivalent to the following,
which we call the \emph{Bapat--Sunder eigenvalue conjecture}.

\begin{conjecture}[Bapat--Sunder \cite{BapatSunder1986}]\label{conj:BS}
If $A$ is an $n\times n$ positive semidefinite Hermitian matrix, then
$\lambda_1(F_A)\le\per(A)$, where $\lambda_1$ denotes the largest
eigenvalue.
\end{conjecture}

In 2018 Drury \cite{Drury2018} disproved the Bapat--Sunder eigenvalue
conjecture with a complex rank-two $8\times 8$ Gram matrix for which
$\lambda_1(F_A)>\per(A)$. Because Drury's witness has non-real
entries, his example did not decide whether the property might still hold for
real symmetric matrices. Drury made this remaining possibility
explicit in the following conjecture:

\begin{conjecture}[Drury {\cite[Conjecture~3]{Drury2018}}]\label{conj:D3}
If $A$ is a real positive semidefinite matrix, then
$\lambda_1(B)=\per(A)$. Here $B=F_A$ is defined by \eqref{eq:FA} and
$\lambda_1(B)$ is its largest eigenvalue.
\end{conjecture}

The restriction to real matrices is natural for two reasons. Before the present work,
the known counterexamples had non-real entries
\cite{Drury2018,Pioge2025}, while the conjecture was known to hold for
rank-one positive semidefinite matrices over either field and for
entrywise nonnegative positive semidefinite matrices
\cite[Corollaries~1 and~2]{Pioge2025}, the latter result being
due to Pate \cite[Theorem~1]{Pate2008}. These results therefore did not
settle Conjecture~\ref{conj:D3}.

The real case also has a physical interpretation. For time-delayed wave
packets with Gaussian envelopes, the relevant Gram matrices are real.
Pioge, Novo and Cerf \cite[\S IV.2]{PiogeNovoCerf2026} observe that
Conjecture~\ref{conj:D3} would rule out anomalous bunching induced by
time delays in interferometers governed by orthogonal matrices.
Theorem~\ref{thm:main} provides such a real violation, and
Corollary~\ref{cor:orthogonal} turns it into a constructive $36$-mode
orthogonal realization. For broader context on permanent conjectures
and their quantum applications, see
\cite{Wanless2022,Zhang2016,Shchesnovich2016PRL,Seron2023,
PiogeSeron2023,PiogeUnified2026}.
In a lossy three-photon interferometer, Ikuta~\cite{Ikuta2026} showed that
conditioning on the survival of all three photons can produce anomalous
generalized bunching even when the unconditioned probabilities satisfy the
Bapat--Sunder Hadamard-product bound.

\begin{theorem}\label{thm:main}
Conjecture~\ref{conj:D3} is false. There is a real positive
semidefinite $32\times 32$ integer matrix $A$ of rank $4$ and an
integer vector $x\in\Z^{32}$ with
\[
x^{\mathsf T}F_A\,x>\per(A)\,x^{\mathsf T}x,
\qquad\text{hence}\qquad
\lambda_1(F_A)>\per(A).
\]
\end{theorem}

Together, Drury's complex counterexample and Theorem~\ref{thm:main}
complete the negative resolution of the Bapat--Sunder eigenvalue
conjecture: it fails for both complex Hermitian and real symmetric
positive semidefinite matrices. Thus restricting the field of scalars
does not recover the conjecture.

Since the inequality in Theorem~\ref{thm:main} is strict and both
$\per$ and $A\mapsto F_A$ are continuous, $A+\varepsilon I$ is a
\emph{positive definite} real counterexample for all sufficiently small
$\varepsilon>0$. Combined with an argument of \cite{Pioge2025},
Theorem~\ref{thm:main} also yields the first counterexample to the
1985 Bapat--Sunder Hadamard-product conjecture \cite{BapatSunder1985}
in which both factors are real and the
second has nonnegative entries
(Section~\ref{sec:consequences}).

\section{The counterexample}\label{sec:cex}

Let $V\in\Z^{4\times 32}$ be the matrix $V=\bigl(V_1\ V_2\bigr)$ with
\begin{equation}\label{eq:V}
\setlength{\arraycolsep}{3.1pt}
\begin{aligned}
V_1&=\begin{pmatrix}
10&1&-6&8&8&9&8&-8&-3&-1&-1&5&-9&6&-4&0\\
-2&6&-6&-5&4&2&2&6&7&-9&-1&1&3&-2&5&3\\
1&-6&-4&-4&4&-2&1&-2&6&3&-8&8&-3&8&4&-1\\
0&5&1&1&-2&4&5&1&4&-2&-5&-1&0&0&-7&9
\end{pmatrix},\\[2pt]
V_2&=\begin{pmatrix}
-1&1&0&-4&8&4&2&0&0&-2&-4&-5&-5&6&7&7\\
-3&3&0&3&5&3&4&8&10&2&7&8&7&-2&-1&7\\
-3&6&7&-3&1&-7&9&6&1&-4&5&3&0&-2&-6&-1\\
-9&7&-7&8&4&5&-2&0&0&9&-3&1&5&7&4&1
\end{pmatrix},
\end{aligned}
\end{equation}
and set $A=V^{\mathsf T}V$. Then $A$ is a real positive semidefinite
integer matrix. The minor of $V$ on its first four columns has
determinant $1005\neq 0$, so $\operatorname{rank}(V)=4$. Since
$\ker(V^{\mathsf T}V)=\ker V$, it follows that $\operatorname{rank}(A)=4$. Let
\begin{equation}\label{eq:x}
\begin{aligned}
x=(&-10,\,24,\,-18,\,0,\,-23,\,1,\,-2,\,-6,\,-3,\,10,\,-15,\,-35,\,
-14,\,-34,\,10,\,17,\\
&\phantom{-}6,\,-4,\,38,\,31,\,-3,\,22,\,-31,\,-10,\,-1,\,31,\,-1,\,
-1,\,8,\,7,\,9,\,-2)^{\mathsf T}.
\end{aligned}
\end{equation}
Exact integer computation (Section~\ref{sec:verif}) gives
\begin{equation}\label{eq:perA}
\begin{aligned}
\per(A)=\ &315675682412132147032972523966676056713438\\
&47182220558452532932009114174163517440000,
\end{aligned}
\end{equation}
$x^{\mathsf T}x=10043$, and
\begin{equation}\label{eq:xBx}
\begin{aligned}
x^{\mathsf T}F_A\,x=\ &31711306091166615356757141988695965627987356\\
&4084711504168249765228465523806473027584000,
\end{aligned}
\end{equation}
so that
\begin{equation}\label{eq:gap}
\begin{aligned}
x^{\mathsf T}F_A\,x-\per(A)\,x^{\mathsf T}x
=\ &799730651618383023571140672268925225673068\\
&33670435629461529060931872682267377664000\ >0.
\end{aligned}
\end{equation}
By the variational characterization of the largest eigenvalue,
\[
\lambda_1(F_A)\;\ge\;\frac{x^{\mathsf T}F_A\,x}{x^{\mathsf T}x}
\;=\;1.000252\ldots\times\per(A)\;>\;\per(A),
\]
which proves Theorem~\ref{thm:main}. A numerical evaluation gives
$\lambda_1(F_A)/\per(A)\approx 1.0004244$.

Two remarks extend the example. Since \eqref{eq:gap} is a strict
polynomial inequality in the entries of $V$, every matrix in a
sufficiently small neighbourhood of $V$ in $\R^{4\times 32}$ is the
Gram factor of a rank-$4$ counterexample. Moreover, counterexamples exist of every order
$n\ge 32$: for $A_m=A\oplus I_m$ the mixed entries of $F_{A_m}$
vanish with $a_{i,j}$, while $\per(A_m(i,i))=\per(A)$ for $i$ in the
identity block, so $F_{A_m}=F_A\oplus\per(A)\,I_m$ and
$\per(A_m)=\per(A)$; hence $\lambda_1(F_{A_m})>\per(A_m)$, and
$A_m+\varepsilon I$ is a positive definite counterexample of order
$32+m$ for all sufficiently small $\varepsilon>0$.

The witness was found by computer search, using an evaluator for
permanents of fixed-rank Gram matrices based on Lemma~\ref{lem:cb},
which makes real configurations of this size tractable.
Realifications and phase/metric deformations of the known complex
counterexamples \cite{Drury2018,Pioge2025} were tried first and
produced no violation, though they motivated searching rank-four real
frames ($\mathbb{C}^{2}\simeq\R^{4}$). The successful configuration
arose independently, from a random Gaussian frame of $50$ unit
vectors in $\R^{4}$ which, scaled by $10$ and rounded to integers,
violates the inequality with Rayleigh ratio approximately $1.033$; greedy
column deletion followed by a fixed-cardinality subset search reduced
it to the $32$ columns displayed. In particular, the witness is not a
realification of the $16$-column complex example of \cite{Pioge2025}:
among its $\binom{32}{2}$ pairs of columns, exactly one is an
equal-norm orthogonal pair, whereas a doubled realification of $16$
complex vectors admits a partition into $16$ such pairs. The parent
frame, its exact certificate, and the reduction to the columns
displayed are included in \cite{dataset}.

The behavior under one- and two-column deletion can also be certified
exactly. For an
$m\times m$ matrix $A'$ obtained by deleting columns of $V$, put
$p=\per(A')$ and let $J$ be the all-ones matrix. Exact evaluation gives
$p>0$ for every $A'$ considered below. If $r,s$ are positive
integers, then positive definiteness of the integer matrix
\[
C_{r,s}=rm p\,I-smF_{A'}+spJ
\]
is equivalent to every eigenvalue of $F_{A'}$ on the orthogonal
complement of the all-ones vector being strictly less than
$(r/s)p$. Fraction-free Bareiss elimination, applied to $C_{123,125}$,
verifies that all its leading principal minors are positive for each of
the $32$ single-column deletions. Thus every resulting $31\times31$
matrix $A'$ satisfies $\lambda_1(F_{A'})=\per(A')$, with every competing
eigenvalue below $0.984\,\per(A')$. Applying the same exact test to
$C_{1,1}$ verifies that none of the $496$ matrices obtained by deleting
two columns violates the inequality either. These computations probe
only subsets of the present witness, and the minimal order of a real
counterexample remains unknown:
Conjecture~\ref{conj:BS} holds for all Hermitian $A$ of order at most
$3$ \cite{BapatSunder1986}, while the present example shows that the
minimal real order is at most $32$.

\section{Exact evaluation of the certificate}\label{sec:verif}

All quantities in \eqref{eq:perA}--\eqref{eq:gap} are integers, and are
evaluated exactly by means of the following classical identity
expressing permanents of Gram matrices in symmetric tensor coordinates.
We include the short proof for completeness. For $\alpha\in\mathbb{N}^{r}$ with
$|\alpha|=\sum_a\alpha_a$, write
$t^{\alpha}=t_1^{\alpha_1}\cdots t_r^{\alpha_r}$ and
$\alpha!=\alpha_1!\cdots\alpha_r!$.

\begin{lemma}\label{lem:cb}
Let $u_1,\dots,u_n,w_1,\dots,w_n\in\R^{r}$, and let $G$ be the
$n\times n$ matrix with $g_{i,j}=u_i\cdot w_j$. Expand
\[
\prod_{k=1}^{n}(u_k\cdot t)=\sum_{|\alpha|=n}c_\alpha\,t^{\alpha},
\qquad
\prod_{k=1}^{n}(w_k\cdot t)=\sum_{|\alpha|=n}d_\alpha\,t^{\alpha},
\qquad t\in\R^{r}.
\]
Then $\per(G)=\sum_{|\alpha|=n}\alpha!\,c_\alpha d_\alpha$.
\end{lemma}

\begin{proof}
Apply the $n$-th order differential operator
$D=\prod_{k=1}^{n}(w_k\cdot\nabla_t)$ to
$p(t)=\prod_{k=1}^{n}(u_k\cdot t)$ and evaluate at $t=0$. Since $p$ is
homogeneous of degree $n$ and each factor of $D$ is a first-order
operator, the Leibniz rule leaves only those terms in which each factor
$w_k\cdot\nabla_t$ differentiates a distinct linear factor of $p$:
terms in which two operators act on the same linear factor vanish
identically, and terms in which some linear factor is left
undifferentiated vanish at $t=0$. Hence
$Dp(0)=\sum_{\sigma\in S_n}\prod_{k=1}^{n}(u_{\sigma(k)}\cdot w_k)
=\per(G)$. On the other hand, applying $D$ termwise,
$D\,t^{\alpha}\big|_{t=0}=\alpha!\,d_\alpha$ whenever $|\alpha|=n$, so
$Dp(0)=\sum_{|\alpha|=n}\alpha!\,c_\alpha d_\alpha$.
\end{proof}

With $v_1,\dots,v_{32}\in\Z^{4}$ the columns of $V$, Lemma~\ref{lem:cb}
applied to $u_k=w_k=v_k$ gives
$\per(A)=\sum_{|\alpha|=32}\alpha!\,c_\alpha^2$, and applied to the
column sets omitting $v_i$ and $v_j$ respectively gives
\[
\per\bigl(A(i,j)\bigr)=\sum_{|\beta|=31}\beta!\,q_{i,\beta}\,q_{j,\beta},
\qquad\text{where}\quad
\prod_{k\neq i}(v_k\cdot t)=\sum_{|\beta|=31}q_{i,\beta}\,t^{\beta}.
\]
Writing $a_{i,j}=\sum_{a=1}^{4}V_{a,i}V_{a,j}$ and combining, one
obtains
\begin{equation}\label{eq:xBxformula}
x^{\mathsf T}F_A\,x=\sum_{a=1}^{4}\ \sum_{|\beta|=31}\beta!
\Bigl(\sum_{i=1}^{32}x_iV_{a,i}\,q_{i,\beta}\Bigr)^{\!2}.
\end{equation}
The right-hand side of \eqref{eq:xBxformula} is a sum of
nonnegative terms, in accordance with the positive semidefiniteness of
$F_A$.

As an independent check on \eqref{eq:perA}--\eqref{eq:gap}, we formed
the matrix $F_A$ explicitly and verified that each of its $32$ row sums equals
$\per(A)$, as the Laplace expansion requires. The witness and the code
performing these computations are available at \cite{dataset}.

\section{Consequences}\label{sec:consequences}

We record four consequences of Theorem~\ref{thm:main}.

\subsection{Algebraic consequences}

By \cite[\S3]{Drury2018}, the spectrum of $F_A$ consists of $\per(A)$
together with that of the block of the Schur matrix of $A$ indexed by
the Ferrers diagram $(n-1,1)$. Question~1 of \cite{Drury2018} asks for
which Ferrers diagrams the largest eigenvalue of the corresponding
block is bounded above by $\per(A)$ for every positive definite
Hermitian $A$; as noted there, for diagrams containing $(7,1)$ ---
among them $(n-1,1)$ for $n\ge 8$ --- a negative answer follows from
Drury's example and the branching rule. Theorem~\ref{thm:main} shows
that the $(n-1,1)$ block can exceed $\per(A)$ already for \emph{real}
$A$, so the analogue of Question~1 in which $A$ ranges over real
positive definite matrices also has a negative answer at
$(n-1,1)$, here with $n=32$, after replacing $A$ by
$A+\varepsilon I$ for sufficiently small $\varepsilon>0$. For which other Ferrers diagrams the
real analogue fails appears to be open.

Two further algebraic consequences concern work of Pioge and
coauthors. Conjecture~4
of \cite{Pioge2025} asserts that
\[
\max\{v^{\mathsf T}F_Av : v\in\R^{n},\ \|v\|_2=1\}\le\per(A)
\]
for all
Hermitian positive semidefinite $A$; it was disproved there by a
complex $16\times 16$ example. For real $A$ the two statements coincide,
so Theorem~\ref{thm:main} shows that Conjecture~4 of \cite{Pioge2025}
already fails for real $A$.

The Bapat--Sunder Hadamard-product conjecture \cite{BapatSunder1985}
asserts that
\[
\per(A\circ B)\le \per(A)\prod_i b_{i,i}
\]
for positive semidefinite Hermitian matrices $A$ and $B$.
Theorem~\ref{thm:main} implies the existence of a counterexample in
which \emph{both} factors are real and the second has nonnegative
entries. The argument is that of
\cite[Corollary~3]{Pioge2025}, with $A$ now real. Let
$v=x/\|x\|_2$ and, for $\varepsilon>0$, let
$B(\varepsilon)=M(\varepsilon)^{\mathsf T}M(\varepsilon)$ be the real
correlation matrix built from $v$ as in \cite[Eq.~(9)]{Pioge2025},
whose entries are
$b_{i,j}(\varepsilon)=(1+\varepsilon^2v_iv_j)/(\alpha_i\alpha_j)$ with
$\alpha_i=\sqrt{1+\varepsilon^2v_i^2}$. Since $v$ is a unit vector
with more than one nonzero coordinate, $\max_i v_i^2<1$, and the
entries of $B(\varepsilon)$ are nonnegative for all
$\varepsilon\le 1$. By \cite[Lemma~1]{Pioge2025},
\[
\per\bigl(A\circ B(\varepsilon)\bigr)
=\per(A)+\varepsilon^{2}\bigl(v^{\mathsf T}F_Av-\per(A)\bigr)
+O(\varepsilon^{4}),
\]
and $v^{\mathsf T}F_Av-\per(A)>0$ by \eqref{eq:gap}. Hence, for all
sufficiently small $\varepsilon>0$,
\[
\per\bigl(A\circ B(\varepsilon)\bigr)>\per(A)
=\per(A)\prod_{i=1}^{32}b_{i,i}(\varepsilon).
\]
The last equality holds because $b_{i,i}(\varepsilon)=1$ for every
$i$. Both $A$ and $B(\varepsilon)$ are real positive semidefinite, and
$B(\varepsilon)$ is entrywise nonnegative.

\begin{corollary}\label{cor:hadamard}
There exist real positive semidefinite matrices $A$ and $B$, with $B$
entrywise nonnegative, such that
$\per(A\circ B)>\per(A)\prod_i b_{i,i}$. In particular, Conjecture~3 of
\cite{Pioge2025} fails over the reals.
\end{corollary}

Consequently, complex phases and negative entries in the second factor
are unnecessary for failure.

As observed in \cite{Pioge2025}, none of the counterexamples to the
Bapat--Sunder Hadamard-product conjecture known before that work
\cite{Drury2016,Drury2017,Seron2023} has a second factor with
nonnegative entries; the counterexample obtained there does, but its
first factor is complex. Note also that by
\cite[Proposition~2]{Pioge2025} (due to Zhang \cite{Zhang1989}), no
counterexample can have an entrywise nonnegative \emph{first} factor.

\subsection{Orthogonal-interferometer realization}

The real Rayleigh-quotient violation has a direct interpretation
for boson bunching with temporal distinguishability. A lossless
interferometer with a real mode transformation is represented by an
orthogonal matrix $U$, whose columns correspond to input modes and
whose rows correspond to output modes. Let $I$ be the occupied input
modes and let $S$ be a designated set of output modes. The matrix
relevant to the probability that all input photons are detected within
$S$ is
\[
H=U_{S,I}^{\mathsf T}U_{S,I}.
\]
Thus $H$ is the Gram matrix of the occupied input columns after they
are restricted to the designated output rows. The criterion of
\cite[\S IV.1]{PiogeNovoCerf2026} shows that a vector $\tau\perp e$,
where $e$ is the all-ones vector, satisfying
\[
\tau^{\mathsf T}F_H\tau>\per(H)
\]
gives a direction of relative arrival-time delays that locally
increases this probability. We show that a positive multiple of the
counterexample in Theorem~\ref{thm:main} can be realized as such a
matrix $H$.

\begin{corollary}\label{cor:orthogonal}
There is a real $36\times36$ orthogonal interferometer with an input of
$32$ pure single photons whose temporal wave packets are Gaussian with
identical variance, and there is a direction of relative arrival-time
delays along which every sufficiently small nonzero delay increases the
probability that all photons are detected within a designated set of
four output modes.
\end{corollary}

\begin{proof}
Write $A=V^{\mathsf T}V$, where $V\in\R^{4\times32}$. To obtain an
orthogonal interferometer, we first embed a scaled copy of $V$ into a
matrix with orthonormal rows. Since
\[
\lambda_{\max}(VV^{\mathsf T})
\leq \operatorname{tr}(VV^{\mathsf T})=3188,
\]
the choice $\gamma=1/6376$ gives
$\gamma VV^{\mathsf T}\preceq\tfrac12 I_4$. Hence the matrix
\[
D=(I_4-\gamma VV^{\mathsf T})^{1/2}
\]
is well defined. The identity $D^2=I_4-\gamma VV^{\mathsf T}$ completes
the row Gram matrix of $\sqrt{\gamma}V$. Indeed, if
\[
R=\bigl(\sqrt{\gamma}\,V\ \ D\bigr)\in\R^{4\times36},
\]
then
\[
RR^{\mathsf T}=\gamma VV^{\mathsf T}+D^2=I_4.
\]
The four rows of $R$ can therefore be completed to a real
$36\times36$ orthogonal matrix $U$.

Take the first $32$ modes as the occupied inputs and the first four
modes as the designated outputs. In the notation of
\cite[Eq.~(3)]{PiogeNovoCerf2026}, the corresponding matrix is
\[
H=U_{S,I}^{\mathsf T}U_{S,I}
=(\sqrt{\gamma}V)^{\mathsf T}(\sqrt{\gamma}V)
=\gamma A.
\]
Equivalently, $H$ is the $32\times32$ principal submatrix of a
rank-four orthogonal projection.

It remains to convert the witness vector $x$ into a vector of relative
arrival-time delays. Adding the same constant to every arrival time has
no physical effect, so the delay vector must be orthogonal to the
all-ones vector. Let $e$ be the all-ones vector. Since
$e^{\mathsf T}x=1$, set
\[
y=x-\tfrac1{32}e,
\qquad \tau=\frac{y}{\|y\|_2}.
\]
Then $e^{\mathsf T}y=0$, so $\tau$ describes relative delays with zero
mean.

To see that this centering does not remove the violation, put
\[
M=F_A-\per(A)I.
\]
The row-sum identity gives $Me=0$. Since $M$ is real symmetric, the
terms involving $e$ vanish, and therefore
\[
y^{\mathsf T}My=x^{\mathsf T}Mx>0
\]
by \eqref{eq:gap}.

Scaling $A$ to $H=\gamma A$ also preserves the sign of the violation.
Since $A$ has order $32$,
\[
F_H=\gamma^{32}F_A,
\qquad \per(H)=\gamma^{32}\per(A).
\]
Consequently,
\[
\tau^{\mathsf T}F_H\tau-\per(H)
=\frac{\gamma^{32}}{y^{\mathsf T}y}
y^{\mathsf T}\bigl(F_A-\per(A)I\bigr)y>0.
\]

Let $\sigma^2$ be the common temporal variance, and let $P(s)$ be the
probability that all $32$ photons are detected within the four
designated output modes when their relative arrival-time vector is
$2\sigma s\tau$. For pure Gaussian wave packets,
\cite[Eq.~(39) and Appendix~C, Eq.~(C6)]{PiogeNovoCerf2026} gives
\[
P'(0)=0,
\qquad
P''(0)=4\bigl(\tau^{\mathsf T}F_H\tau-\per(H)\bigr)>0.
\]
Hence
\[
P(s)=P(0)+\tfrac12P''(0)s^2+o(s^2)>P(0)
\]
for all sufficiently small nonzero $s$. Thus the direction $\tau$ has
the asserted local effect.
\end{proof}

\section{Concluding remarks}

The Bapat--Sunder eigenvalue conjecture is false over both the
complex and real fields. Drury's counterexample established failure
for complex Hermitian matrices; Theorem~\ref{thm:main} eliminates the
remaining possibility that reality restores the property. By
Corollary~\ref{cor:hadamard}, the Bapat--Sunder Hadamard-product inequality can fail for real
factors with the second entrywise nonnegative. For the eigenvalue
inequality, rank-one positive semidefinite matrices and entrywise
nonnegative positive semidefinite matrices remain classes on which it
holds \cite[Corollaries~1 and~2]{Pioge2025}.
Corollary~\ref{cor:orthogonal} shows that, for identical-variance
Gaussian wave packets, a direction of small arrival-time delays can
produce anomalous photon bunching in a real orthogonal interferometer,
here with all $32$ photons detected within four designated output modes
of a $36$-mode device.
Also open are the
minimal rank of a real counterexample, where rank two would bear on
Drury's remaining conjecture \cite[Conjecture~2]{Drury2018}, and the
question of which other blocks of the Schur matrix can exceed $\per(A)$
for real $A$.

\section*{Disclosure of generative AI use}

OpenAI's GPT-5.6 Sol and Anthropic's Claude Fable 5 assisted with
implementation of the search described in Section~\ref{sec:cex}; the
latter also assisted with drafting. The author takes full
responsibility for the content of the article.

\section*{Disclosure statement}

The author reports there are no competing interests to declare.

\section*{Funding}

This research did not receive any specific grant from funding agencies
in the public, commercial, or not-for-profit sectors.

\section*{Data availability statement}

The witness matrices, exact certificates and verification code are
openly available on Zenodo at
\url{https://doi.org/10.5281/zenodo.21543697}.

\end{document}